\documentclass[11pt,a4paper]{article}

\usepackage{natbib}

\usepackage{amsmath,amsthm,bm}\usepackage{mathrsfs}
\usepackage{amsfonts}
\usepackage{textcomp}
\usepackage{upgreek}
\usepackage{indentfirst}
\usepackage{booktabs}
\usepackage{threeparttable}
\usepackage{multirow}
\usepackage{authblk}
\usepackage[dvipsnames,rgb,dvips]{xcolor}
\usepackage{graphicx}
\usepackage{geometry}
\usepackage{caption}
\usepackage{overpic}
\usepackage{enumitem}
\theoremstyle{definition}

\theoremstyle{remark}

\numberwithin{equation}{section}
\newcommand{\ve}[1]{\ensuremath{\mbox{\boldmath$#1$}}}
\newcommand{\ma}[1]{\ensuremath{\mathbb{#1}}}

\begin{document}

\title{Navigation of micro-swimmers in steady flow: the importance of symmetries}%
\author[1]{Jingran Qiu}
\author[2]{Navid Mousavi}
\author[2]{Kristian Gustavsson}
\author[1]{Chunxiao Xu}
\author[2]{Bernhard Mehlig}
\author[1]{Lihao Zhao\thanks{corresponding author}}
\affil[1]{AML, Department of Engineering Mechanics, Tsinghua University, 100084 Beijing, China}
\affil[2]{Department of Physics, University of Gothenburg, SE-41296 Gothenburg, Sweden}

\date{\today}%

\maketitle

\begin{abstract}
Marine microorganisms must cope with complex flow patterns and even turbulence as they navigate the ocean. To survive they must avoid predation and find efficient energy sources. A major difficulty in analysing possible survival strategies is that the time series of environmental cues in non-linear flow is complex, and that it depends on the decisions taken by the organism. One way of determining and evaluating optimal strategies is reinforcement learning. In a proof-of-principle study, Colabrese {\em et al.} [Phys. Rev. Lett. (2017)] 
used this method to find out how a micro-swimmer in a vortex flow can navigate towards the surface as quickly as possible, given a fixed swimming speed.  The swimmer measured its instantaneous swimming direction and the local flow vorticity in the laboratory frame,
and reacted to these cues by swimming either left, right, up, or down. However, usually a motile microorganism measures the local flow rather
than global information, and it can only react in relation to the local flow, because in general it cannot access global information (such as up or down in the laboratory frame). Here we analyse optimal strategies with local signals and actions that do not refer to the laboratory frame. We demonstrate that symmetry-breaking is required in order to learn vertical migration in a meaningful way. Using reinforcement learning we analyse the emerging strategies for different sets of environmental cues that microorganisms are known to measure.
\end{abstract}

\section{Introduction}
\label{sec1}
Active microorganisms  are ubiquitous in marine environments. They use mechanoreceptors to measure hydrodynamic signals, and they can adjust their motion by rotating to change their swimming direction, or by accelerating vigorously \citep{Kiorboe1999hydro}. This allows planktonic copepods, for example, to evolve advanced swimming strategies.
They can maintain their depth in the ocean by swimming against upwelling or downwelling flows~\citep{Genin2005}. In order to maintain efficient swimming~\citep{Michalec2015,Michalec2017} or to avoid predation \citep{GemmellAdhikari2014}, copepods can also adapt their swimming strategy in response to a changing turbulence intensity.

How did these strategies evolve? This question requires answers on different levels. From an evolutionary perspective, what is the cost or reward function to be optimised? Is it more important to avoid predation, or to find food in an efficient way? To answer such questions starting from a mechanistic model for a micro-swimmer in a complex flow poses several challenges, not least concerning the  fluid mechanics of small swimming organisms.

First of all, which signals can an active microorganism measure, and how? This is quite well understood \citep{VisserBook}.
Planktonic copepods, for example, use arrays of setae to detect small velocity differences to the ambient flow. This mechanism is very sensitive. Velocity differences as small as 20 \textmu m/s can be measured reliably \citep{Yen1992}. From the bending patterns of their setae, these organisms  can also distinguish their relative angular velocity to the fluid, as well as its strain rate~\citep{Kiorboe1999hydro}.   

Second, given a reward function to optimise, which environmental signals are the most important?
Third, given certain environmental cues, what should the swimmer do? It should act rationally \citep{VisserBook}, because the strategies evolved under natural selection must benefit the survival of the organism. Different species developed different strategies. Copepods tend to jump when they sense a fluid disturbance caused by a predator \citep{Kiorboe1999hydro}, or they may adjust their cruising speeds slowly in response to a changing turbulence intensity \citep{Michalec2015}. Sometimes, a copepod may change not only
its swimming speed but also its swimming direction \citep{kiorboe2010unsteady}.
When oyster larvae sense a change in turbulent intensity close to the seafloor, they dive and attach to the floor~\citep{Fuchs2013}. Phytoplankton can modulate the efficiency of vertical migration \citep{Durham2013,Gustavsson2016} in response to turbulence, by changing cellular morphology \citep{Sengupta2017} or forming cell chains \citep{Park2001,Lovecchio2019}.
Vertical migration is a fundamental phenomenon for marine microorganisms, related to nutrition uptake, reproduction, and predation~\citep{Huntley1982,Hays1994,Katajisto1998,Park2001}.
Many microorganisms are gyrotactic to promote vertical migration.
Such swimmers have an inhomogeneous mass density so that gravity tends to align them head up, causing them to swim upwards  \citep{Kes85,Kessler1986individual}.

Fourth, if one tracks a microorganism in a  complex or turbulent flow, how should one interpret its actions? The challenge is that the cues may change in an apparently random fashion as the organism explores the flow. In general it is difficult to model the time sequence of environmental cues, because it depends on the actions the swimmer chooses to take. In addition, what looks like a good move at any moment may turn out not to be optimal in the long run.

In summary, the question is how to find optimal strategies to optimise a given reward for a  micro-swimmer in a complex flow, and how to understand the  mechanisms that determine the optimal strategy. In a recent proof-of-principle study, \citet{Colabrese2017} demonstrated that reinforcement learning is an efficient way to
address this question. The authors used the Q-learning algorithm~\citep{Watkins1992,Sutton1998,Mehlig2021}
to investigate how a gyrotactic micro-swimmer can learn efficient vertical migration.
The point is that fluid-velocity gradients, tend to re-orient the swimmer, making it difficult to find the optimal path to the surface.
\citet{Colabrese2017} analysed different swimming strategies for an idealised model of a swimmer in a two-dimensional steady vortex flow.
The swimmer measured the local vorticity of the flow, whether it was positive, negative, or close to zero,
and whether the current swimming direction pointed left, right, up, or down in the laboratory frame. The possible actions was to swim left, right, up, or down.
\citet{Colabrese2017} demonstrated that smart swimmers can avoid being trapped in vortices, and that they can take advantage of upwelling flows to accelerate upward navigation. This work motivated some follow-up studies \citep{Gustavsson2017,Colabrese2018,Alageshan2020,Qiu2020}, investigating different flows, as well as different actions.
In all of these studies, some signals and actions referred to the laboratory frame, so that the swimmer had, in effect, access to a map, which facilitated navigation. The navigation problems considered in recent studies \citep{biferale2019chaos,schneider2019optimal,muinos2021reinforcement,gunnarson2021learning} also used information relating to a fixed reference frame.

Motile microorganisms in a complex flow, by contrast, do not carry a map. They can only access
limited information regarding the local flow field. By detecting the velocity difference to the surrounding flow, an organism can estimate the local fluid-velocity gradients or its rotation  relative to that of the local fluid. One way of inferring these signals is through bending patterns of setae  \citep{Kiorboe1999hydro,Kiorboe1999prey,VisserBook}. In some cases, a swimmer might have access to global information. During day time, it can for instance follow the light to find the way to the surface (phototaxis) \citep{Cohen2002}. This does not work at night or under conditions where it is hard to determine the light direction.
Therefore, the question is  how a micro-swimmer in complex flow can learn to navigate, given that it has only local information in its frame of reference. The answer to this question depends on the reward function, and in particular on its symmetries. We expect that a task that does not break any symmetry of the problem is easiest to learn, such as escaping from a certain point as far as possible in a given time with a given swimming speed. Efficient upward migration, by contrast, may be more difficult to learn because it breaks rotational symmetry.


In this paper we use reinforcement learning to find efficient strategies for vertical migration using only local signals and local actions. We use a highly idealised model for a motile microorganism and its hydrodynamic sensing capabilities. With Q-learning, we search for optimal strategies for fast vertical migration in a two-dimensional steady Taylor-Green vortex flow \citep{taylor1923viii}, which allows for direct comparison with \citet{Colabrese2017}, and in a steady two-dimensional random velocity field. We investigate how symmetry breaking  allows the swimmer to distinguish different directions (which is necessary to swim upwards), highlighting significant differences between navigation using local signals in the frame of reference of the swimmer and signals in the laboratory frame~\citep{Colabrese2017,Colabrese2018,Gustavsson2017,Alageshan2020,biferale2019chaos,schneider2019optimal,muinos2021reinforcement,gunnarson2021learning}. We find that symmetry-breaking due to gravity allows the swimmers to find efficient strategies for vertical migration. To achieve this, they emulate more slender swimmers through adaptive steering, which allows them to preferentially sample upwelling regions of the flow to accelerate upward migration.

\section{Methods}
\label{sec2}

\subsection{Model}
\label{model}
\begin{figure}
	
	\includegraphics[width = 8cm]{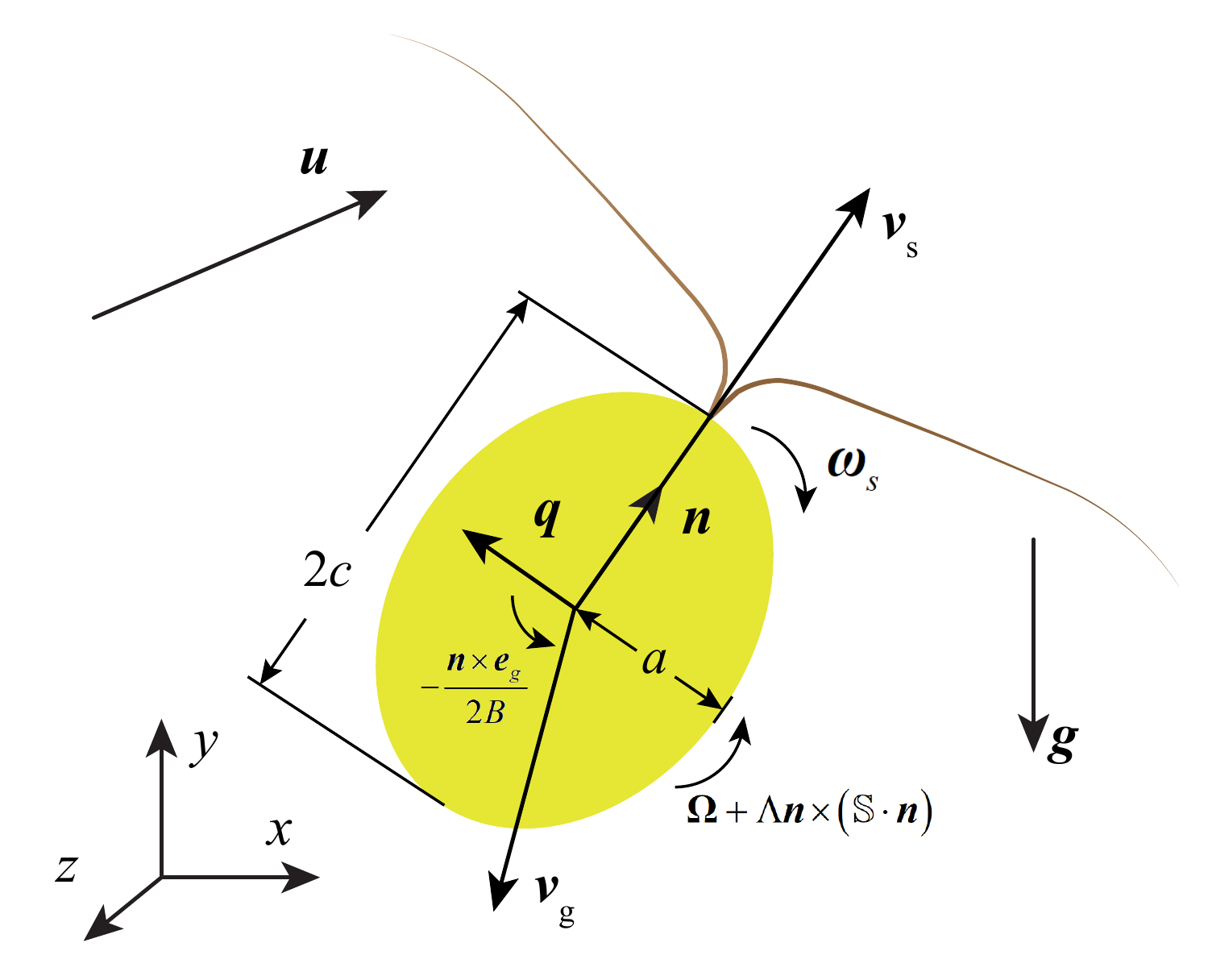}
	\centering
	\caption{Sketch of an elongated micro-swimmer in the  $x$-$y$ plane, showing the velocities and angular velocities that determine the dynamics of the micro-swimmer [Eq.~(\ref{eom})].
	}
	\label{figsketch}
	
\end{figure}

We consider an active swimmer that cruises with a constant translational speed $v_{\rm s}$ and can choose to rotate with an angular velocity $\ve \omega_s$.
We model the swimmer as an elongated spheroid with aspect ratio $\lambda=c/a$ and symmetry axis $\ve n$ (Figure~\ref{figsketch}). Using typical parameter values for motile microorganisms in the ocean (Table~\ref{tab1}), we infer that the Reynolds number, ${\rm Re}_{\rm p}=2cv_{\rm s}/\nu\sim 0.2$, and the Stokes number, $\mathrm{St} = \tau_p / \tau_f \sim 0.003$, are both small. Here, $\nu$ is the kinematic viscosity, $\tau_{p}$ is the particle response time for spheroids \citep{Kim:2005,zhao2015rotation}, and $\tau_f$ is a characteristic flow time-scale (discussed in more detail below). Assuming that $\rm Re_p\ll 1$ and $\rm St\ll 1$, we neglect the inertia of swimmer and fluid, and use the following overdamped model
\citep{Durham2013,Gustavsson2016}  for the dynamics
\begin{subequations}
	\label{eom}
	\begin{align}
		\label{xdot}\dot{\ve{x}}&=\ve{v}\,,\\
		\label{vdot}\ve{v}&=\ve{u}+v_{\rm s}\ve{n}+\ve{v}_{\rm g}+\ve{\xi}\,,
		\\
		\label{ndot}
		\dot{\ve n}&=\ve{\omega} \times \ve{n}\,,\\
		\ve{\omega}&=\ve{\Omega}
		+ \Lambda \ve{n}\times\left( \ma{S}\ve{n}\right)
		+ \ve{\omega}_{\rm s}
		-\tfrac{1}{2B}{\ve{n}\times \ve{e}_{\rm g}}+\ve{\eta}\,.
		\label{pdot}
	\end{align}
Here, $\ve v$  and $\ve\omega$ denote the velocity and the angular velocity of the swimmer, respectively. The first three terms on the right-hand-side (r.h.s.) of Eq. (\ref{vdot}) denote the fluid velocity $\ve u(\ve x,t)$ at the swimmer position $\ve x$, the constant translational swimming speed $v_{\rm s}$ in the direction of $\ve n$, and the Stokes settling velocity $\ve v_{\rm g}$ for a spheroid~\citep{Kim:2005}:
	\begin{equation}
		\label{vsettle}
		\ve v_{\rm g}=v_{\rm g}^\perp \ve{e}_{\rm g}
		+ \left[ v_{\rm g}^{\parallel} -v_{\rm g}^\perp\right]
		\left( \ve{e}_{\rm g} \cdot \ve{n}\right) \ve{n}\,.
	\end{equation}
\end{subequations}
Here, $\ve{e}_{\rm g}$ is the unit vector in the direction of gravity, and $v_{\rm g}^\parallel$ and $v_{\rm g}^\perp$ are the settling speed of a spheroid in a quiescent fluid aligned parallel with or perpendicular to gravity, respectively \citep{Kim:2005}. The first two terms on the r.h.s. of Eq.~(\ref{pdot}) correspond to Jeffery's angular velocity for a spheroid, using the Stokes approximation~\citep{Jef22}. These contributions to the angular velocity
depend on the flow vorticity $2\ve\Omega$, the strain $\ma S$, and the shape factor
\begin{equation}
	\Lambda = \frac{\lambda^2-1}{\lambda^2+1}\,.
	\label{eqLambda}
\end{equation}
The third term on the r.h.s. of Eq. (\ref{pdot}) represents the  angular velocity due to active rotational swimming, $\ve\omega_{\rm s}$. The fourth term
is the gyrotactic angular velocity due to a restoring torque towards $-\ve{e}_{\rm g}$, with a time scale $B$ \citep{Kes85,VisserBook,Durham2013,Gustavsson2016}.
Finally, $\ve{\xi}$ and $\ve \eta$ in Eqs. (\ref{eom}) represent small white-noise perturbations, added to remove the influence of initial position and orientation of the swimmers, and to break structurally unstable periodic orbits \citep{Colabrese2017}.

In the present study we consider only two-dimensional flows. Although Eqs. (\ref{eom}) are compatible with three-dimensional motion, we restrict translational and rotational motion of the swimmer to the $x$-$y$ plane.
We use two different models for the fluid-velocity field. First, to compare with \citet{Colabrese2017}, we consider a two-dimensional, time-independent Taylor-Green vortex (TGV) flow  with velocity \citep{taylor1923viii}:
\begin{equation}
	\label{TGV}
	\ve u(\ve x)=\ve\nabla\times\ve e_z \psi(\ve x)\quad\mbox{with stream function}\quad \psi(\ve x)=-\frac{u_0L_0}{2}\cos\frac{x}{L_0}\cos\frac{y}{L_0}\,.
\end{equation}
Here $\ve e_z$ is the unit vector in the $z$-direction and we assume that the gravitational acceleration points in the negative $y$-direction, $\ve e_{\rm g}=-\ve e_y$ as shown in Figure \ref{figsketch}.
We take $u_0=2.0$~mm/s and $L_0 =0.5$~mm for the velocity and length scales of the flow, and define $\tau_f \equiv L_0/u_0$. These scales correspond to the Kolmogorov scales \citep{Fri97} of ocean turbulence with kinematic viscosity $\nu\sim 10^{-6}\text{m}^2/s$ and energy dissipation rate $\mathscr{E}=1.6\times 10^{-5}\text{m}^2\text{s}^{-3}$~\citep{Yamazaki1996}. Following the values in Table \ref{tab1} and the assumed velocity and length scales of the flow, the non-dimensional parameters of swimming velocity and gyrotactic stability \citep{Durham2011,Colabrese2017} are
\begin{equation}
	\Phi \equiv \frac{v_{\rm s}}{u_0} = 0.5\quad \mbox{and}\quad \Psi\equiv \frac{BL_0}{u_0}=20\,.
\end{equation}
Comparing with \citet{Durham2011} and \citet{Colabrese2017}, who explored a parameter range of $0.01 <\Phi,\Psi <100$, our swimmers have low swimming speeds and weak gyrotaxis. Therefore, they cannot migrate efficiently unless they actively adjust their swimming directions.

To verify the generality of the results obtained for the TGV flow, we compare with results obtained using a Gaussian random velocity field. This velocity field is defined by a stream function $\psi(\ve x)$ with zero mean and correlation function (see \citet{Gus15a} for details)
\begin{align}
	\label{Stochastic}
	\langle\psi(\ve x)\psi(\ve x')\rangle=\frac{\ell^2u_{\rm s,0}^2}{2}
	\exp\left[-\frac{x^2}{2\ell^2}\right]\,.
\end{align}
We choose the parameters $\ell$ and $u_{\rm s,0}$ so that the spatial averages of $\ve u^2$ and $\sum_{i,j}(\partial_iu_j)^2$ are equal those obtained from Eq.~(\ref{TGV}).

\begin{table}	
	\begin{center}
	\begin{tabular}{llccc}\hline\hline
		&& range & value used & unit\\
				\hline
		Swimmer size  & $2c$& 0.1-0.5& 0.2&mm \\
		Aspect ratio & 2.0-2.5& 2.0& \\
		Mass-density ratio & $\rho_p/\rho_f$& 1.005-1.019& 1.017& \\
		Swimming velocity& $v_{\rm s}$ & 0.33-3.76& 1.00& mm/s \\
		\multirow{2}*{Settling velocity} & $v_{\rm g}^\parallel $&\multirow{2}*{0.1-0.8}& 0.152& mm/s\\
		~& $v_{\rm g}^\perp $&~& 0.133& mm/s\\
		Swimming angular velocity & $\omega_{\rm s}$ & $<$20.0 & 1.0 & rad/s \\
		Gyrotactic timescale & $B$& $\approx$ 5.0 & 5.0& s\\
				\hline\hline	
	\end{tabular}
	\caption{
		Summary of model parameters. The swimmer size is obtained from the length of a small copepod \citep{Titelman2001,Titelman2003} and the aspect ratio is a rough estimate between the length and width of copepods~\citep{Carlotti2007}. The mass-density ratio is calculated using mass density of copepods $\rho_p$ \citep{Knutsen2001} and a sea-water density of $\rho_f = 1.025\mathrm{g}/\mathrm{cm}^3$ (salinity of 3.5\% and temperature of $20^\circ\mathrm{C}$) \citep{millero1980new}. The value of the swimming velocity is typical of values observed in experiments \citep{Titelman2003}. The settling velocity is estimated by Stokes settling velocity \citep{Kim:2005}, which is consistent with the range given in \citet{Titelman2003}. The maximal angular velocity can be estimated from the images shown in \citet{jiang2004} to about $90^\circ$ in 0.067s. We use a lower value for convenience in numerical simulation, representing the slow steering motion \citep{kabata1971locomotory}.
		The gyrotactic re-orientation time is taken from \citet{Fields1997} for copepods.	
	}	
	\label{tab1}
	\end{center}
\end{table}

\subsection{Hydrodynamic signals}
\label{secsignals}
A small microorganism can sense the local motion of the surrounding fluid with its setae by detecting velocity differences between its body and the fluid \citep{VisserBook,Kiorboe1999hydro,Kiorboe1999prey}. For example, to first approximation a small microorganism can be considered rigid, so that it cannot deform.
If the surrounding fluid deforms with strain rate $\ma S$, velocity differences $\ve \delta_{\rm s}$ must arise between the surface velocity of the swimmer
and the fluid velocity: $\ve \delta_{\rm s} = \ma S\ve r$, where $\ve r$ is the vector from the centre of a swimmer to a point on its surface \citep{Kiorboe1999prey}. In this way, a swimmer can detect the fluid strain rate in its local frame of reference. This argument neglects the fact that the swimmer disturbs the flow. Strictly speaking, it can therefore not directly measure the undisturbed fluid strain while swimming.
Accurate measurement of the strain rate is difficult when the swimming speed is of the same order of magnitude as the fluid velocity~\citep{VisserBook}. However, it is nevertheless plausible that the swimmer can encode the signal of the undisturbed flow by many densely distributed sensors along the body~\citep{Fields2014}.

At any rate, for an incompressible velocity field in two dimensions there are two independent strain parameters. Assuming that the swimmer can measure the normal and tangential components of the fluid strain rate tensor along its swimming direction, we take the independent components to be
\begin{subequations}
	\label{signals}
	\begin{align}
		\label{I}
		S_{nn}&=\ve{n}\cdot \ma{S} \ve{n}\,,\\
		\label{J}
		S_{nq}&=\ve{n}\cdot \ma{S} \ve{q}\,.
	\end{align}
	Here $\ve n$ is the swimming direction defined above, and $\ve q$ is a  vector orthogonal to $\ve n$, such that $\ve n$ and $\ve q$ form a right-handed orthonormal basis in the flow plane (Figure \ref{figsketch}).
	
	Relative rotation between the  fluid and the swimmer also results in velocity differences on the surface of the swimmer, given by $\ve \delta_\Omega \sim (\ve \Omega-\ve \omega)\times \ve r$, where both $\ve\Omega$ and $\ve\omega$ are parallel to $\ve e_z$ in a two-dimensional flow. Relative angular motion results from the gyrotactic torque \citep{VisserBook}, or simply because the swimmer rotates actively. We assume that the swimmer can measure local relative rotation:
	\begin{align}
		\label{K}
		\Delta\Omega&= (\ve \Omega-\ve \omega)\cdot \ve e_z\,.
	\end{align}
	
Finally, microorganisms can also measure the local slip velocity, due to fluid acceleration, swimming, or settling~\citep{VisserBook}. In two spatial dimensions,there are two independent components of the slip velocity:
	\begin{align}
		\label{L}
		\Delta u_{n}&= \left(\ve{u}-\ve{v}\right) \cdot \ve{n}\,,\\
		\label{M}
		\Delta u_{q}&= \left(\ve{u}-\ve{v}\right) \cdot \ve{q}\,.
	\end{align}
\end{subequations}

\begin{table}
	\begin{center}
	\begin{tabular}{llccc}\hline\hline
		signal &&  & threshold &\\
		\hline
		strain rate  & $S_{nn},S_{nq}$& & $S_{\rm c}=0.5$&
		$\text{s}^{-1}$\\
		angular slip velocity & $\Delta\Omega$& & $\Delta\Omega_{\rm c}=0.5$& $\text{s}^{-1}$\\	
		slip velocity& $\Delta u_q$& & $\Delta u_{\rm c}=50$& \textmu m/s\\   
		\hline\hline	
	\end{tabular}
	\caption{
	Summary of signals and thresholds we use in Q-learning. 
	The threshold values $S_{\rm c}$, $\Delta \Omega_{\rm c}$, and $\Delta u_{\rm c}$ are used to discretise the signals for Q-learning. The value of $S_{\rm c}$ is taken from experiments where copepods are observed to jump in response to strain rates above $\sim 0.5s^{-1}$~\citep{Kiorboe1999hydro,buskey2002escape}. The values of $\Delta \Omega_{\rm c}$ and $\Delta u_{\rm c}$ are then estimated from $S_{\rm c}$, see text.}
	\label{tab:thresholds}
	\end{center}
\end{table}

In order to implement the Q-learning algorithm, we must discretise the signals. Appropriate discretisation scales can be estimated from the threshold of sensing velocity differences $\Delta u_{\rm c}$, and from the size $c$ of the swimmer~\citep{Kiorboe1999prey}.
This results in the following scales for the thresholds of strain $S_{\rm c}= \Delta u_{\rm c} /c$ and angular slip velocity, $\Delta\Omega_{\rm c}= \Delta u_{\rm c} / c$.
Thresholds estimated using the half length of the major axis $c$ (instead of the minor axis $a$) reflect the highest sensitivity to signals for spheroidal swimmers: any signal below these thresholds cannot give rise to a velocity difference greater than $\Delta u_{\rm c}$ anywhere on the swimmer surface. 
In experiments it is observed that copepods make escape jumps in response to strain rates with threshold values in a range of more than one order of magnitude \citep{Kiorboe1999hydro,buskey2002escape}. Here we adopt a typical value, $S_{\rm c}\sim 0.5s^{-1}$\,, corresponding to $\Delta u_{\rm c}= cS_{\rm c}\sim 50$ \textmu m/s, which is of the same order of magnitude as the smallest velocity difference, 20~\textmu m/s, that a swimmer can physically measure~\citep{Yen1992}.
From this value we also obtain $\Delta\Omega_{\rm c}= S_{\rm c} = 0.5~\text{s}^{-1}$ from the definition above.
The signal $\Delta u_{n}$ evaluated using the swimming speed in Table \ref{tab1} lies below the lower threshold, $\Delta u_{n} < -\Delta u_{\rm c}$. This signal is therefore always activated and the swimmer can therefore not distinguish changes of the signal $\Delta u_{n}$ close to the threshold value. Rather than introducing new, arbitrary thresholds for $\Delta u_{n}$, we focus on the other four signals in Eqs.~(\ref{signals}) in what follows, on  $S_{nn}$, $S_{nq}$, $\Delta \Omega$ and $\Delta u_q$.
In Table~\ref{tab:thresholds} we summarise the signals and thresholds used.

\subsection{States and actions}
\label{secaction}
To apply the Q-learning algorithm in its simplest form, we must define states and actions.
The state of the swimmer is obtained from local measurements of the environment.
Given the thresholds quoted above, we discretise each of the signals in Table~\ref{tab:thresholds} into three states. For example, the possible values of $S_{nn}$ are discretised into three intervals $S_{nn}<-S_{\rm c}$,  $-S_{\rm c} < S_{nn}<S_{\rm c}$, and  $S_{nn}>S_{\rm c}$.

Depending on the state of swimmer, it may take different actions.
In our model, the swimmer moves with constant speed, while steering with an angular velocity $\ve\omega_{\rm s}$~\citep{kabata1971locomotory}.
For two-dimensional flows, only the $z$-component, $\omega_{\rm s}\equiv \ve\omega_{\rm s}\cdot\ve e_z$ matters.
We assume that $\omega_{\rm s}$ can take three values, $\{-1,0,1\}$ rad/s. These are the actions the swimmer can take: either it swims straight ahead, $\omega_{\rm s}=0$, or it steers with constant positive or negative angular velocity. We choose $1$ rad/s for the magnitude of the steering angular velocity, which is one quarter of the maximum flow vorticity, because this allows the swimmer to exert some orientational control in regions of lower vorticity. This value is by a factor of ten smaller than the angular velocity that would be obtained if the swimmer were to convert its full propulsion effort into angular motion, $\omega_{\rm max}\sim v_{\rm s}/c \sim 10$ rad/s. This means that the steering motion only requires a small amount of energy compared to that required for propulsion.
We also remark that some copepods can achieve much higher angular velocities, up to about 20~rad/s, when they rotate rapidly before they jump  \citep{jiang2004}.
Our model does not describe such vigorous motion.

These states and actions are local. They refer to the frame of reference of the swimmer, and do not directly relate to the laboratory frame. We contrast this with the states and actions stipulated by \citet{Colabrese2017}. They defined the states of the swimmer in terms of discrete orientations in the laboratory frame (left, right, up, or down), and the three discretised levels of the vorticity of background flow ($\Omega < -\Omega_{\rm c}$, $-\Omega_{\rm c} < \Omega < \Omega_{\rm c}$ and $\Omega > \Omega_{\rm c}$, with the threshold $\Omega_{\rm c} = u_0/6L_0$). The actions of the swimmer in \citet{Colabrese2017} are to rotate with angular velocity
\begin{equation}
	\label{eq:naive}
	\ve \omega_{\rm s} = \frac{1}{2B}(\ve n \times \ve k)\,.
\end{equation}
The vector $\ve k$ is chosen by the swimmer from four possible directions (left, right, up, or down) in the laboratory frame in the $x$-$y$ plane. Below we compare strategies obtained for both models.

\subsection{Q-learning}
\label{secQL}
The task of the swimmer is to navigate upward through the flow. As mentioned in the Introduction, vertical migration is common and important for
microorganisms in the ocean. Since this task breaks rotational symmetry, it allows us to illustrate the role of symmetries in the learning problem.
In order to find optimal upward navigation strategies, we use the reward function
\begin{equation}
	\label{eq:reward}
	r_i=\frac{1}{L_0}\left(y_{i+1}-y_i\right)\,.
\end{equation}
Here $y_i$ is the vertical location of the swimmer immediately after a state update $s_{i-1}\to s_i$. For the simulation in TGV flow, states are updated at a fixed time step size, and $r_i$ is thus proportional to the time averaged velocity from $s_{i}$ to $s_{i+1}$. This allows the algorithm to optimise the vertical navigation velocity. For the simulation in random velocity fields, states are updated only when one of the state signals changes its discretised state level, and $r_i$ is only approximately proportional to a velocity (see Appendix~\ref{appendix}). We have confirmed that these two update rules give the same result in TGV flow.

We use the one-step Q-learning algorithm~\citep{Watkins1992,Sutton1998,Mehlig2021} to search for efficient strategies for vertical migration. The swimmers move according to the dynamics (\ref{eom}) with $\omega_{\rm s}$ adjusted according to the current state. When evaluating strategies we use a greedy choice of action: whenever the state is updated to $s_i$, the swimmer takes the action $a_i = \arg\max_{a} Q(s_i,a)$. The value function $Q(s_i,a)$ is an estimate of the summation of future reward if action $a$ is taken at state $s_i$, also referred to as Q table. To find an estimate of the Q table, we use a training phase, where the swimmer adopts the $\varepsilon$-greedy strategy: it mainly takes the action $a_i = \arg\max_{a} Q(s_i,a)$, but takes a random action with a probability $\varepsilon$. This allows the swimmer to explore different actions and helps to avoid local optima. Given $\left\lbrace s_i,a_i,r_{i},s_{i+1}\right\rbrace $, the Q table is updated in the standard fashion during the training phase:
\begin{equation}
	\label{updateQ}
	Q\left( s_i,a_i\right) \gets Q\left( s_i,a_i\right)+
	\alpha \left[ r_{i}+ \gamma \mathop{\max}_{\alpha} Q\left( s_{i+1},a\right)
	-Q\left( s_i,a_i\right) \right].
\end{equation}
The learning rate $\alpha$ is a free parameter to control the convergence speed. 
The discount rate $\gamma$, $0\le \gamma<1$, is introduced to regularise the value of the Q table in the long run. We choose $\gamma=0.999$ to obtain a far-sighted strategy. 
The training is divided into episodes. In each episode, ten swimmers sharing the same Q table are initialised with random locations and orientations.
Then the swimmers navigate the flow for a fixed amount of state changes corresponding to a physical time, $T_N=250$ s, after which a new one begins. The Q table converges to an approximately optimal policy after a sufficient number of episodes. Further details concerning the training and the corresponding parameter values are given in Appendix \ref{appendix}.

\begin{table}
	\centering	
	\begin{tabular}{ccccccc}
		\hline\hline
		\multirow{2}*{case}&\multicolumn{3}{c}{symmetry-breaking information} &\multirow{2}*{training} \\
		~&states &actions &dynamics \\
		\hline
		S1& no& no& no&  failed \\
		S2& yes& no& no& succeeded \\
		S3& no& yes& no&  succeeded \\
		S4& no& no& gyrotaxis&  succeeded \\
		S5&  no& no& settling&  succeeded\\
		S6& no& no& gyrotaxis \& settling&  succeeded\\
		\hline\hline
	\end{tabular}
	\caption{Cases studied (Section \ref{secsym}).}
	\label{tabsym}
\end{table}

\subsection{Summary of cases studied}
\label{secsym}

As mentioned in the Introduction, our goal is to investigate the role of symmetries in finding optimal strategies for vertical migration.
Table \ref{tabsym} summarises the different cases we analyse, S1 to S6. The TGV flow has $C_4$ point-group symmetry. If neither signals, nor actions, nor the dynamics
break this symmetry, the swimmer cannot learn to distinguish the $\ve e_y$-direction. It cannot learn to navigate in this direction (case S1). For cases S2 and S3, either signals or actions break this symmetry, similar to the cases studied by \citet{Colabrese2017}. Since the swimmer has direct access to the laboratory coordinates, it learns to navigate as expected.
For cases S4 and S5,  neither states nor actions break the $C_4$ symmetry. In this case the swimmer can learn to migrate along the $\ve e_y$-direction if its dynamics breaks the symmetry, either because it experiences a gyrotactic torque, or because the swimmer is heavier than the fluid and settles along the negative $\ve e_y$-direction. These cases are analysed in Section \ref{sec:sb}. 

It is clear that gyrotaxis in case S4 must help the swimmer to learn successfully, since it tends to align $\ve n$ with the $\ve e_y$-direction. But  training can be successful
even in the absence of gyrotaxis.  It might appear that settling alone should make it more difficult for the swimmer in case S5 to navigate upwards, because settling imposes a negative contribution to $v_y$ after all. But in the absence of any other symmetry breaking, settling may enable the swimmer to move upwards although gravity pulls it down. 

Finally for case S6, both settling and gyrotaxis act with the parameters given in Table \ref{tab1}. This case is analysed in Section \ref{subsec3.2}, to understand the microscopic mechanisms that allow the swimmer to learn with local signals and actions.

\section{Results}
\label{sec3}

\subsection{Symmetry breaking}
\label{sec:sb}
\begin{figure}	
	\centering
	\begin{overpic}[width=6.5cm]{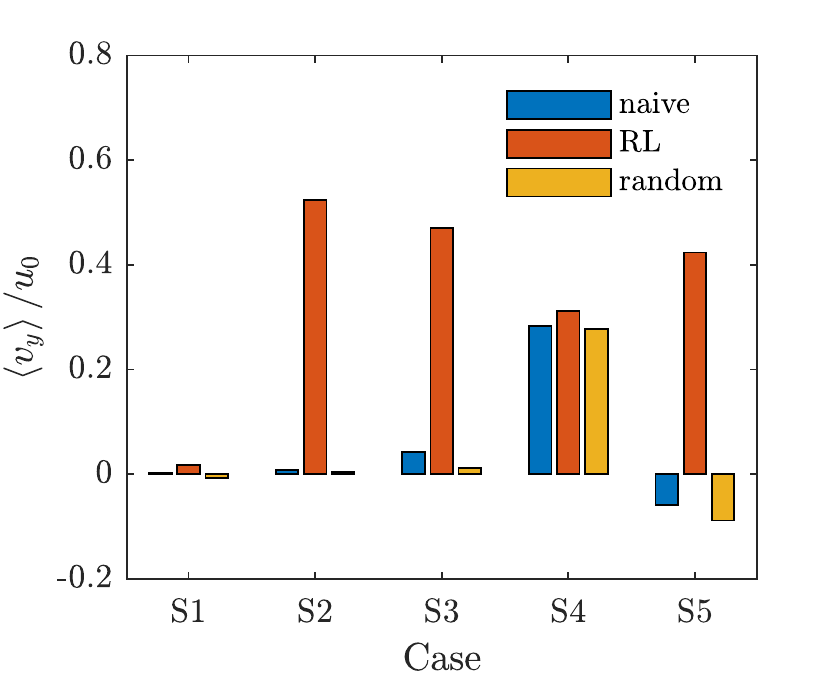}
		\put(50,82){\colorbox{white}{\phantom{XXXX}}}
	\end{overpic}
	\captionof{figure}{Normalised averaged vertical velocity  $\langle v_y\rangle$ for cases S1 to S5 (Table \ref{tab1}). Shown are the results following the naive strategy (\lq{}naive\rq{}, see text), the best strategy obtained by reinforcement learning (\lq{}RL\rq{}) using the signals $S_{nn}$ and $S_{nq}$, and a random strategy (\lq{}random\rq{}), where the swimmer
		takes random actions when its state defined by $S_{nn}$ and $S_{nq}$ changes.
	}
	\label{symm_gain}
\end{figure}

To find the optimal strategy with reinforcement learning, we use a subset of signals, only $S_{nn}$ and $S_{nq}$. Each signal gives rise to three states, which means that the size of the Q table is $3^2 \times 3$, corresponding to 9 states and 3 actions. Thus, a swimmer can choose between $3^9 \sim 10^4$ different strategies in total. In principle, one can evaluate the performance of each possible strategy, but Q-learning allows us to obtain optimal or approximately optimal strategies much more efficiently.
Figure \ref{symm_gain} illustrates the training results for cases S1 to S5  (Table \ref{tabsym}). Shown is the average velocity of swimmers $\langle v_y\rangle$ [Eq. (\ref{vdot})] in the $y$-direction after they have reached a steady state. Angular brackets represent the ensemble average over the positions of swimmers. Red bars show the results of the best strategy obtained after training in each case.

To assess the success of the optimal strategy, we compare it
with two others. First, we consider a swimmer that follows the \lq{}naive\rq{} strategy, following a single predefined action. For S3 this means that $\omega_{\rm s}$ is chosen according to Eq.~(\ref{eq:naive}) with $\ve k=\ve e_y$, so that
the swimmer always turns towards the  $\ve e_y$-direction in the laboratory frame \citep{Colabrese2017}. This strategy breaks the $C_4$ symmetry. For cases S2, S4, and S5, the naive strategy corresponds to  $\omega_{\rm s}=0$, which does not break this symmetry. Second, adopting a random strategy, the swimmer chooses a random action with equal probability whenever the state defined by $S_{nn}$ and $S_{nq}$ changes. The random strategy preserves the point-group symmetry, at least on average.

Figure \ref{symm_gain} shows, as expected, that  learning fails in the completely symmetric case S1. The results for cases S2 and S3 confirm, as expected, that the swimmers can learn to optimise vertical migration when either signals or actions break the symmetry of the flow \citep{Colabrese2017}. We also see that the optimal strategy found by reinforcement learning is better than both naive or random strategies, resulting in a larger $\langle v_y\rangle$.
Cases S4 and S5 show that the swimmer can still learn to optimise vertical migration. In both cases, the $C_4$ symmetry is neither broken by signals nor actions, but by the dynamics. With gyrotaxis alone (no settling, case S4), the optimal strategy is only slightly better than the naive one. This is not surprising, because the naive strategy breaks the symmetry.
It is interesting, however,  that settling alone helps the swimmer to navigate upwards (no gyrotaxis, case S5). If the symmetry is not broken in any other way, settling is in fact necessary to allow the swimmer to find the positive $\ve e_y$-direction. We see in Figure \ref{symm_gain} that the signals $S_{nn}$ and $S_{nq}$ provide enough information for the swimmer to actively exploit the flow.
In the next Section we discuss the underlying mechanisms.

\begin{figure}
	\centering
	\includegraphics[width=10cm]{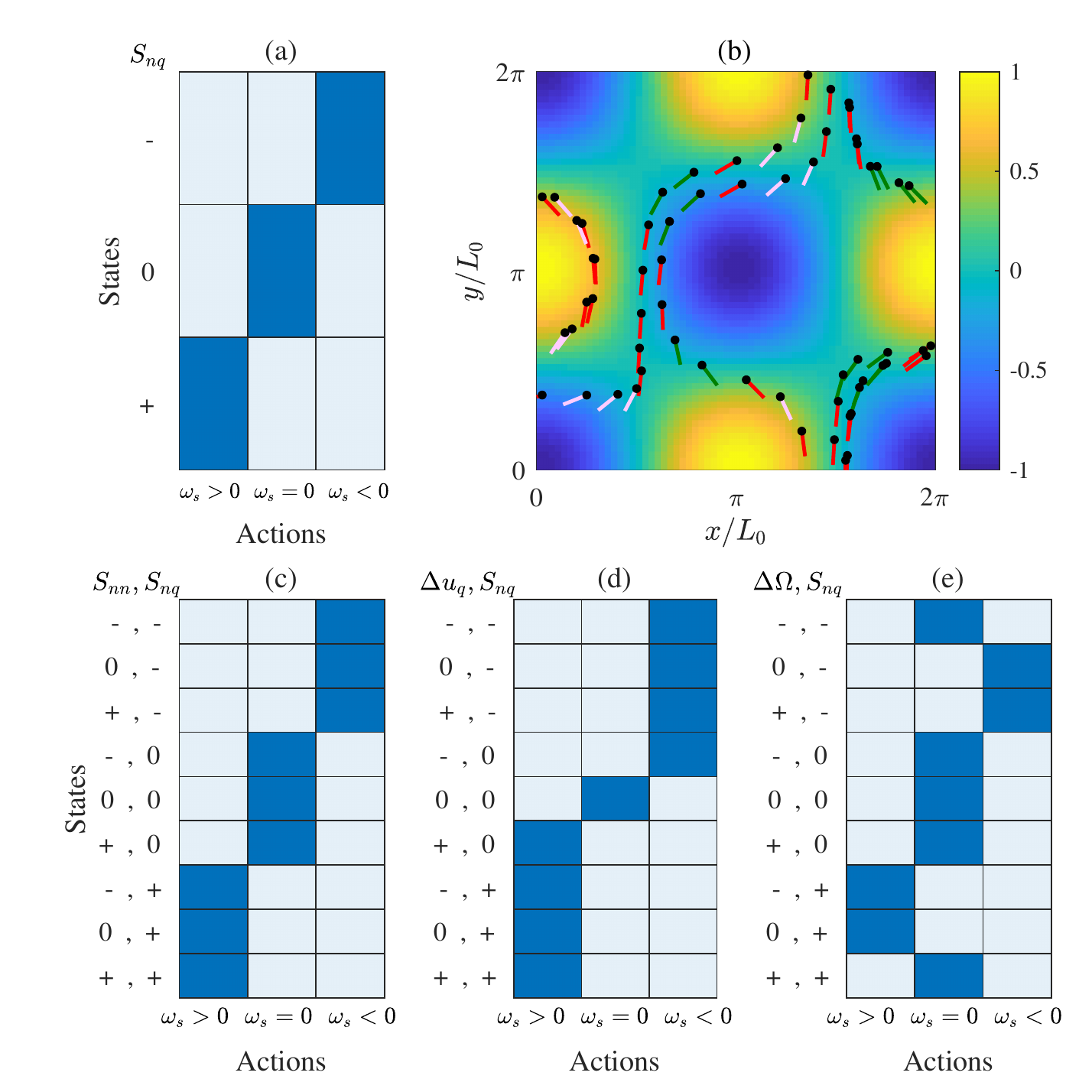}
	\caption{(a) Q table of the optimal strategy when swimmers sense only $S_{nq}$. Each signal has three levels: negative ($-$), approximately zero ($0$), or positive ($+$), as described in Section \ref{secaction}. The cells filled with blue indicate the optimal action for each state. (b) Typical trajectories  of smart swimmers following the strategy shown in panel (a). Black dots represent the instantaneous position of the swimmer, and the coloured line segments indicate the swimmer orientation $\ve n$ (representing the tail of the swimmer).  The colours represent signals and corresponding actions: green: $S_{nq}<0$, $\omega_s<0$; red: $S_{nq}\approx0$, , $\omega_s=0$; salmon: $S_{nq}>0$, $\omega_s>0$. The background colour gives the normalised vorticity $2\Omega_zL_0/u_0$. Remaining panels: Q tables for different signal choices, using  $S_{nn}$-$S_{nq}$ (c),  $\Delta u_q$-$S_{nq}$ (d), and $\Delta\Omega$-$S_{nq}$ (e).
		\label{figQtab}
	}
	\label{compare}
\end{figure}

\subsection{Mechanisms}
\label{subsec3.2}
\label{secMechanisms}
\begin{figure}
	\centering
	\includegraphics[width=\textwidth]{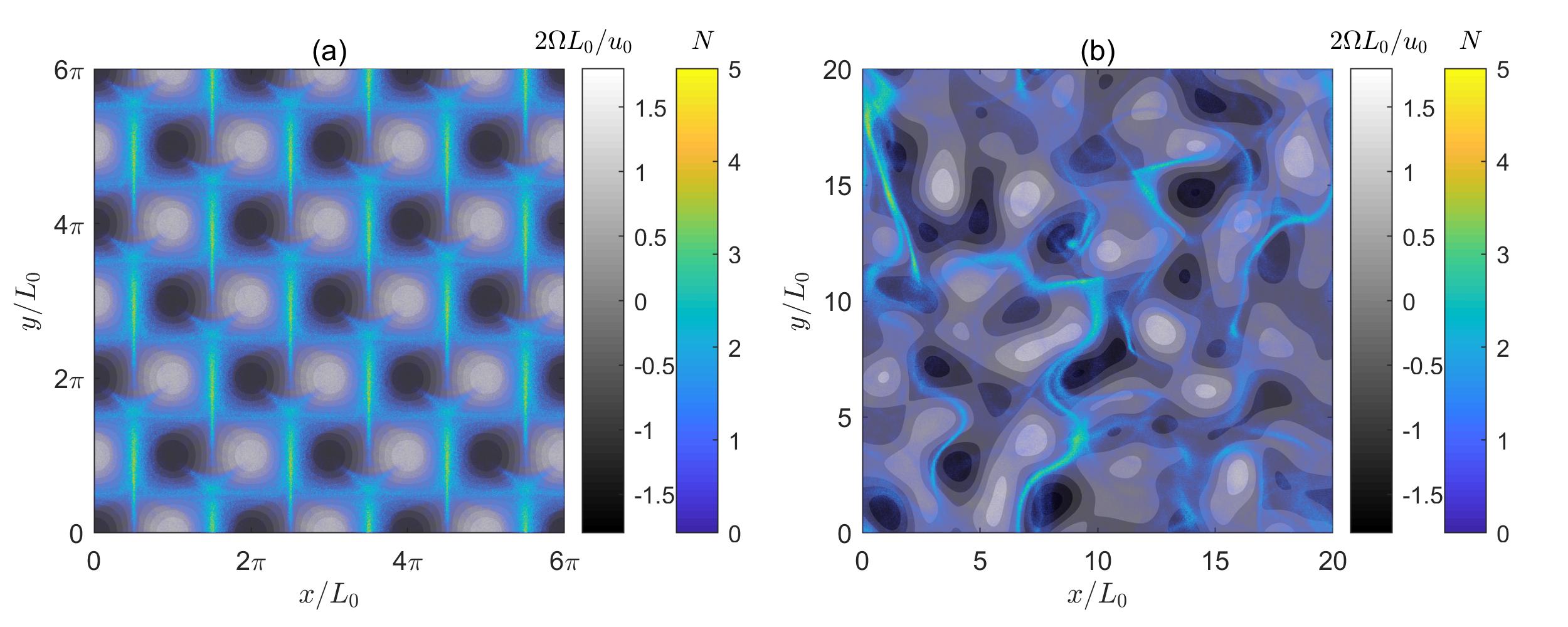}	
	\caption{Number density $N$ of swimmers following the dynamics (\ref{eom}) with angular swimming $\omega_{\rm s}$ according to the strategy in Figure \ref{figQtab}(a) (colour scale) in (a) the TGV flow [Eq.(\ref{TGV})] and in (b) the random velocity field [Eq. (\ref{Stochastic})]. The grey background scale refers to the normalised fluid vorticity, $2\Omega L_0/u_0$.}
	\label{figdensity}
\end{figure}

How does the swimmer make use of local signals to navigate? We consider a swimmer following the dynamics  (\ref{eom}), with the parameters
given in Table \ref{tab1}. The steering angular velocities are $\omega_{\rm s}=-1, 0,1$ rad/s
as described in Section \ref{secaction}, and the signals are taken to be subsets of those listed in Table \ref{tab:thresholds}.
Figure  \ref{figQtab} refers to four different combinations: $S_{nq}$ alone, $S_{nn}$ \& $S_{nq}$,  $\Delta u_q$ \& $S_{nq}$, and $\Delta\Omega$ \& $S_{nq}$.
The key message is that $S_{nq}$ alone allows the swimmer to successfully navigate. The corresponding Q table is shown in
Figure  \ref{figQtab}(a), and typical trajectories of swimmers following this strategy are shown in panel (b). We see that the  swimmer learns to avoid the regions of strong vorticity and finds upwelling regions where the background flow tends to carry it upwards. Figure \ref{figdensity} illustrates that this behaviour is not particular to the TGV flow. Panel (a) shows
how smart swimmers preferentially sample the upwelling fringes of the vortices in the TGV flow: they swim to the right of positive vortices (with $\Omega>0$, white),
and to the left of negative vortices (black). Panel (b) shows the same, but for swimmers navigating a spatially smooth, steady
random velocity field [Eq.~(\ref{Stochastic})]. This suggests that the learned strategy is robust, at least for steady two-dimensional flows, for parameter values similar to those shown in Table~\ref{tab1}.

The underlying strategy in Figure \ref{figQtab}(a) relies entirely on the signal $S_{nq}$. For a swimmer moving in a two-dimensional plane (the $x$-$y$-plane) we have
\begin{equation}
	S_{nq}=\ve{n}\cdot \ma{S}  \ve{q}\ = \ve{e}_z \cdot \left[ \ve n \times \left( \ma S \ve n\right)\right]\,.
	\label{eqsnq}
\end{equation}
Comparing with the equation of motion [Eq. (\ref{pdot})], we see that $S_{nq}$  determines how the strain rotates the swimmer.
When $S_{nq}$ is negative, for example, a prolate swimmer ($\Lambda >0$) is rotated clockwise by the strain. Panels (a) and (b) in Figure~\ref{figQtab} show
that the optimal action does the same:  $\omega_{\rm s}<0$ means that the swimmer steers clockwise. For $S_{nq}>0$, on the other hand, the flow turns
the swimmer counter-clockwise, and so does the optimal action. Finally, when $S_{nq}$ is close to zero, the swimmer does not actively steer.

Since the steering mirrors the effect of the strain, we conclude that the swimmer tries to emulate a more slender swimmer, with a larger value of $\Lambda$.
This makes sense, because it was shown by \citet{Gustavsson2016} that naive gyrotactic swimmers (no steering actions, $\omega_{\rm s}=0$) tend to sample upwelling regions of the flow
when their shape factor $\Lambda$ increases, or at least regions where downwelling is weaker. In other words, the smart swimmers manage to preferentially sample upwelling regions of the flow by mimicking slender naive swimmers that take no steering actions ($\omega_{\rm s}=0$).

\begin{figure}
	\centering
	\includegraphics[width=8cm]{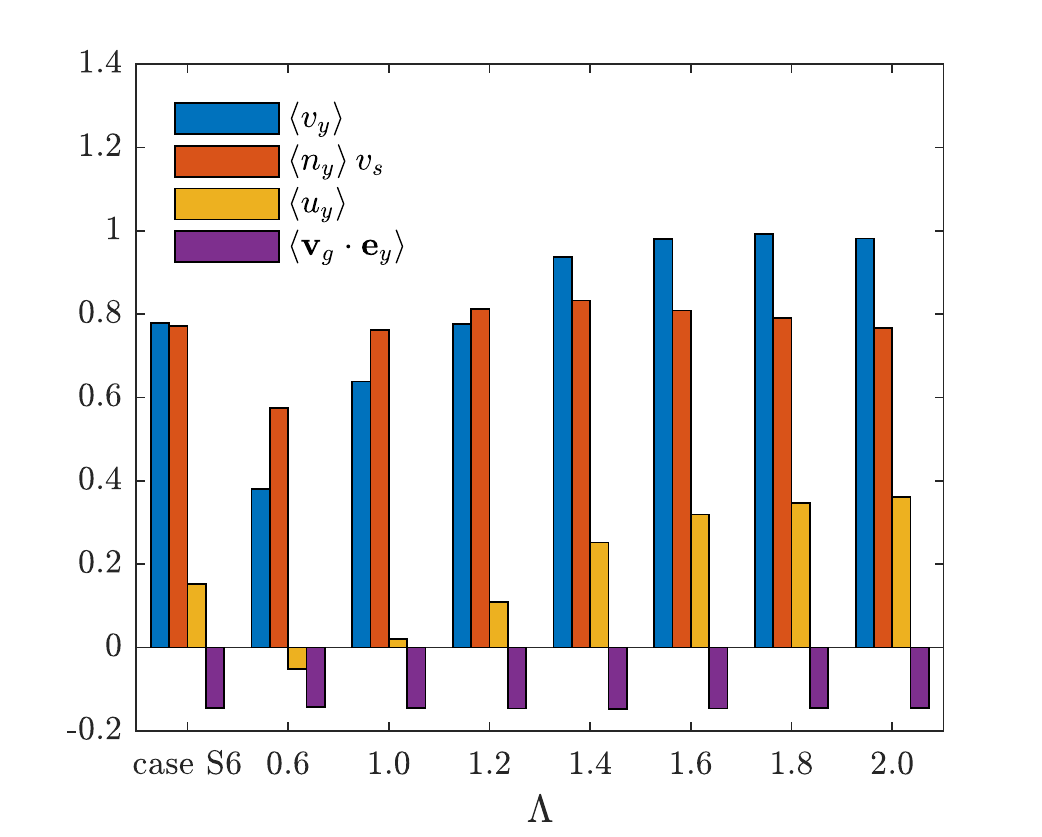}	
	\caption{Contributions to the vertical swimming of swimmers defined in case S6 (the leftmost group) and naive swimmers with different $\Lambda$ (the other groups) in the TGV flow. The blue bars represent the average of vertical velocity $\langle v_y\rangle$. The red, yellow and purple bars represent the vertical velocity due to swimming, advecting of local fluid, and settling, respectively. All velocities are normalised by the swimming velocity $v_{\rm s}$.}
	\label{figLda}
\end{figure}
How this works is shown more quantitatively in Figure \ref{figLda}, where we compare the performance of smart swimmers with $\Lambda =0.6$ to naive gyrotactic swimmers
with $\Lambda \geq 0.6$. Figure \ref{figLda} shows their mean vertical velocities $\langle v_y\rangle$, and also the different contributions to  $\langle v_y\rangle$, namely $\langle n_y\rangle v_{\rm s}$,
$\langle u_y\rangle$, and the average component of the settling velocity in the direction of gravity, $\langle \ve v_{\rm g}\cdot \ve e_{\rm g}\rangle$.
We see that the smart swimmers have an appreciable upward velocity, because gyrotaxis favours alignment between $\ve n$ and $\ve e_y$,
and because the swimmers sample upwelling regions where $u_y >0$. Naive swimmers with the same value of $\Lambda$ also migrate upwards, but more slowly. Figure \ref{figLda} reveals the reason: naive swimmers do not sample upwelling regions as efficiently as smart ones and do not align as much with the upward direction. Figure~\ref{figLda} also illustrates that naive swimmers with larger values of $\Lambda$ tend to sample upwelling regions more efficiently ($\langle u_y\rangle$ is larger) and align more with the upward direction than in the case $\Lambda=0.6$. For a spheroid, the shape factor $\Lambda$ is constrained to be smaller than unity [Eq. (\ref{eqLambda})]. It is nevertheless interesting to investigate the dynamics for larger values of $\Lambda$, artificially increasing the effect of the strain on the angular dynamics \citep{zhao2019passive}. Here, we change the value of $\Lambda$, but keep $\lambda$, $v_g^{\parallel}$ and $v_g^{\perp}$ constant as given in Table \ref{tab1} (this means that Eq. (\ref{eqLambda}) no longer holds).

Figure \ref{figLda} shows that a smart swimmer with shape factor $\Lambda=0.6$ migrates as fast as a naive swimmer with $\Lambda = 1.2$, roughly twice as fast as a naive swimmer of the same parameter $\Lambda=0.6$. We see that the average vertical velocity $\langle v_y\rangle$ increases as $\Lambda$ grows until $\Lambda=2$. This is explained by three observations. First, the average vertical swimming velocity $\langle n_y\rangle v_s$ increases as $\Lambda$ increases from 0.6 to 1.4, because $\ve n$ tends to align with $\ve e_y$. But note that the alignment becomes weaker for $\Lambda>1.4$.
Second, as predicted in \citet{Gustavsson2016} and later verified in direct numerical simulation of turbulence~\citep{borgnino2018gyrotactic,Lovecchio2019,cencini2019gyrotactic}, the dynamics of gyrotactic swimmers undergoes a flow-and-parameter dependent transition from preferential sampling of downwelling regions for $\Lambda<\Lambda_{\rm c}$ to upwelling regions for $\Lambda >\Lambda_{\rm c}$.
For the TGV flow and parameters leading to Figure~\ref{figLda}, this transition occurs at $\Lambda_{\rm c}\approx 1$ as shown in Figure~\ref{figLda}. This implies that naive gyrotactic swimmers cannot sample upwelling regions. 
Smart swimmers, on the other hand, succeed in sampling upwelling regions by emulating naive swimmers with $\Lambda >1$.
Third, the settling velocity  $\langle \ve v_{\rm g}\cdot \ve e_{\rm g}\rangle$ is found to be almost independent of $\Lambda$. This is explained by the observation that the orientation $\ve n$ has only a small influence on the settling velocity since $v_{\rm g}^{\parallel}$ and $v_{\rm g}^\perp$ are of the same order of magnitude for the parameters given in Table~\ref{tab1}.
These three observations explain the strategy shown in Figure \ref{figQtab}(a), as well as the trajectory patterns shown in Figures~\ref{figQtab}(b)~and~\ref{figdensity}.

Up to now we discussed the simplest case, where the swimmers sense only one signal, $S_{nq}$ [Figure \ref{figQtab}(a)]. Now consider the cases where the swimmers sense not only $S_{nq}$ but also other signals.
In each case, we obtain an efficient strategy as shown in panels (c) to (e) of Figure~\ref{figQtab}, respectively. These strategies are robust, because they are learned in both TGV flow and a random velocity field. However, these strategies are only slightly better than the one in Figure \ref{figQtab}(a): the resulting upward velocities are of the same order of magnitude for all the four strategies. Moreover, these strategies share a similar pattern.
For instance, the strategy in Figure~\ref{figQtab}(c) yields actions independent of the signal $S_{nn}$. In other words, it is identical to the strategy obtained using only $S_{nq}$ as a signal [Figure~\ref{figQtab}(a)].
For the cases where the swimmers measure $\Delta u_q$ or $\Delta\Omega$ in addition to $S_{nq}$, the Q tables shown in panels (d) and (e) illustrate that $\Delta u_q$ is useful only when $S_{nq}$ is close to zero, and $\Delta\Omega$ is useful only when $S_{nq}$ and $\Delta\Omega$ are either both positive or negative. This means measuring extra signal in addition to $S_{nq}$ provides not much useful information for the upward navigation of swimmers.
We, therefore, conclude that it is $S_{nq}$ that provides the most important information, because in two-dimensional flows it is a direct measure of the rotation the fluid strain exerts upon the swimmer, and it allows the swimmers to smartly mimic more elongated swimmers.

\section{Conclusions}
\label{sec4}
We analysed how a micro-swimmer can learn to migrate upwards in a complex flow, relying only on local hydrodynamic signals. We assumed that the swimmer can actively rotate in its frame of reference in response to the local signals. Using reinforcement learning, we found that successful swimming strategies exist only if the swimmer can distinguish the target direction. The flows we considered do not provide direct information concerning the target: the TGV flow has $C_4$ rotational symmetry, and the random velocity field is statistically isotropic. We demonstrated that in this case the dynamics must break the symmetry, to allow the swimmer to learn meaningful strategies.  For example, gravity breaks the symmetry of the dynamics by gyrotaxis or settling, and this allows the swimmer to navigate with local signals and actions.
For the case of a gravitational force, we found that even though settling opposes the migration task, the resulting breaking of rotational symmetry allows the swimmer to swim upwards, with an average velocity larger than the settling velocity.

We investigated the underlying mechanism of navigation with local signals by considering swimmers with both gyrotaxis and settling. Reinforcement learning projects out a simple but efficient optimal strategy for vertical navigation:
the swimmer measures the strain component, $S_{nq}$, allowing it to adjust its angular velocity to amplify strain-mediated rotations, leading to a preferential sampling of strain regions and upwelling regions.
For the parameters tested here, we found the same optimal strategy in both the TGV flow and different realisations of a smooth random velocity field, indicating that the strategy is robust. For the tested cases, the strategy leads to twice the navigation speed compared with a naive gyrotactic swimmer~\citep{Kes85,Kessler1986individual,Durham2011,Durham2013,Gustavsson2016}. 

To find efficient survival strategies for microorganisms in the ocean, a number of open questions remain.
In our model we considered steady flows in two spatial dimensions. How will the optimal strategy change in three-dimensional unsteady flows?
Moreover, in our model the perceived signals were given by estimated threshold values, but in reality the swimmers may respond to other levels of hydrodynamic signals. How can we determine which level is the most important? Is it possible to model the signal under disturbances from swimming?

Here we only considered the swimming strategy for vertical migration, given a reward function to optimise the migration rate. To optimise this rate is a fairly simple task, and this choice of target allowed us to study the effect of symmetry breaking, and to compare with the existing literature on navigation of smart swimmers.
However, in nature there are competing tasks, such as minimising energy consumption during migration, or avoiding predation~\citep{Huntley1982,morris1985propulsion,Park2001}. How to analyse the effect of these competing tasks remains an open question, necessary to answer in order to understand observed survival strategies of motile microorganisms in the ocean.

{\em Acknowledgements.} KG and BM are supported in part by  a grant from the Knut and Alice Wallenberg Foundation, grant no. 2019.0079, and in part by VR grant no. 2017-3865.
	KG is supported by Vetenskapsrådet, Grant No. 2018–03974.
	BM and LZ are supported by a collaboration grant from the joint China-Sweden mobility programme [National Natural Science Foundation of China (NSFC)-Swedish Foundation for International Cooperation in Research and Higher Education (STINT)],  grant nos. 11911530141 (NSFC) and CH2018-7737 (STINT). JQ and LZ acknowledge the support from the Institute for Guo Qiang of Tsinghua University (Grant No. 2019GQG1012). CX acknowledges the support from NSFC (Grant No. 91752205).

\appendix
\section{Numerical details}
\label{appendix}

\begin{table}
	\centering	
	\begin{tabular}{llccccc} \hline\hline
		flow field& case&$\alpha_0$ &$\sigma_0$&$\varepsilon_0$ & $E_0$ & Episode number\\
				\hline
		\multirow{9}*{TGV}&S1& 0.02& 1000& 0.0& $\infty$ &1000 \\
		~&S2& 0.02& 1000& 0.0& $\infty$ &1000 \\
		~&S3& 0.02& 1000& 0.0& $\infty$ &1000 \\
		~&S4& 0.02& 1000& 0.0& $\infty$ &1000 \\
		~&S5& 0.02& 1000& 0.0& $\infty$ &1000 \\
		
		~&S6: $S_{nn}$-$S_{nq}$& 0.008& 1000& 0.001& 2000&2000 \\
		~&S6: $S_{nq}$& 0.02& 1000& 0.0005& 1000&1000 \\
		~&S6: $S_{nq}$-$\Delta u_{q}$& 0.02& 1000& 0.0005& 1000&1000 \\
		~&S6: $S_{nq}$-$\Delta\Omega$& 0.01& 1000& 0.001& 2000&2000 \\
		random velocity field &S6: all cases & 0.04& 1000& 0.02& 2000&3000\\
		
		
		\hline
	\end{tabular}
	\caption{Training parameters.}
	\label{taba}
\end{table}

In this appendix we give details for the different setups in Table~\ref{tabsym}.
To implement the Q-learning algorithm both actions and states must be discretised.
The action is to modify the angular swimming velocity $\ve \omega_{\rm s}$ in Eq.~(\ref{pdot}).
In cases S1, S2, S4, S5, and S6, the actions are $\ve\omega_{\rm s}=(0,0,\omega_z)$ with $\omega_z$ taking one of the values $- 1, 0, \text{or} \ 1 \text{ rad}/s$. In case S3 the actions are given by Eq.~(\ref{eq:naive}), with $\ve k$ being one of $(0,-1,0)$, $(1,0,0)$, $(0,1,0)$ or $(-1,0,0)$~\citep{Colabrese2017}.
The states are given by different combinations of the signals.
In cases S1, S3, S4, and S5, the states are given by the combination of $S_{nn}$ and $S_{nq}$.
In case S2, the states are given by the combination of three levels of local vorticity and four discrete levels of the instantaneous direction of the swimmer~\citep{Colabrese2017}.
In case S6, the states are given either by the signal $S_{nq}$ solely, or in combination with one of $S_{nn}$, $\Delta u_q$ and $\Delta \Omega$~[Eqs.~(\ref{signals})].

In each episode, ten swimmers are initialised with random locations and orientations.
For the simulation in TGV, we integrate Eqs. (\ref{eom}) numerically using an explicit second-order Adams-Bashforth scheme for $10^5$ time steps of size $\Delta t = 0.01u_0/L_0$, allowing the swimmers to reach a steady state. The Gaussian noises $\ve \xi$ and $\ve \eta$ in Eqs. (\ref{vdot}) and (\ref{pdot}) are implemented at every time step by Gaussian translational and rotational displacements with variances $\sqrt{2D_t \Delta t}$ and $\sqrt{2D_r \Delta t}$, respectively, where $D_t = 0.001L_0 u_0$ and $D_r = 0.01u_0/L_0$.
The state is updated every 10 time steps, giving a total of $10^4$ state updates each episode. We use $\gamma = 0.999$ to optimise the average velocity on a time window of order $(1-\gamma)^{-1}=10^3$ state changes, corresponding to $25~\text{s}$ in physical time. This time window is much longer than the time scale for a swimmer to move through a vortex, estimated by $L_0/v_{\rm s} = 0.5~\text{s}$, and thus, is sufficient for optimizing the average velocity in a steady state.

During the training, we take both the learning rate, $\alpha$, and the exploration rate, $\varepsilon$, to decrease as the episode number $E$ increases, i.e.
\begin{equation}
	\alpha=\alpha_0\frac{\sigma_0}{\sigma_0+E}, \quad
	\varepsilon=\varepsilon_0 \max\left(0, 1- \frac{E}{E_0} \right)\,,
\end{equation}
with initial learning and exploration rates $\alpha_0$ and $\varepsilon_0$ and decay scales $\sigma_0$ and $E_0$. Training parameters for all cases are shown in Table \ref{taba}.

To verify the robustness of the swimming strategies learned in TGV, we also do the training in random velocity fields. The result shows that, the strategies learned in TGV, as shown in Figure~\ref{figQtab}, also have high performance and can be learned in random velocity fields. 
There are some differences in the implementation of training in random velocity fields. First, a state is updated only when one of the state signals changes discretised state level. Second, each episode lasts for 2500 s, corresponding to $10^6$ time steps. Third, we use the Euler method to integrate Eqs.(\ref{eom}). In principle, these differences in numerical implementation do not have substantial influence on the symmetry problem and the mechanisms of migration with local signals. We have also confirmed that the two approaches give the same result for the TGV flow.


\end{document}